\documentclass[runningheads,a4paper]{llncs}

\usepackage{amssymb}
\usepackage{graphicx}

\usepackage{url}

\newcommand{\keywords}[1]{\par\addvspace\baselineskip
\noindent\keywordname\enspace\ignorespaces#1}

\begin{document}

\mainmatter  

\title{Genetic Algorithm Modeling with GPU Parallel Computing Technology}

\titlerunning{Genetic Algorithm Modeling with GPU Parallel Computing Technology}

%
%
\author{Stefano Cavuoti\inst{1}\and Mauro Garofalo\inst{2}\and Massimo Brescia\inst{3,1}%
\thanks{corresponding author, brescia@oacn.inaf.it}%
\and Antonio Pescape'\inst{2}\and\\
Giuseppe Longo\inst{1,4}\and Giorgio Ventre\inst{2}}
\authorrunning{Cavuoti et al.}

\institute{Department of Physics, University Federico II, Via Cinthia 6, I-80126 Napoli, Italy\\
\and
Department of Computer Engineering and Systems, University Federico II, Via Claudio 21, I-80125 Napoli, Italy\\
\and
INAF, Astronomical Observatory of Capodimonte, Via Moiariello 16, I-80131 Napoli, Italy\\
\and
Visiting Associate, California Institute of Technology, Pasadena, CA 91125, USA}

%
%

\toctitle{Lecture Notes in Computer Science}
\tocauthor{S. Cavuoti et al.}
\maketitle

\begin{abstract}
We present a multi-purpose genetic algorithm, designed and implemented with GPGPU / CUDA parallel computing technology. The model was derived from a multi-core CPU serial implementation, named GAME, already scientifically successfully tested and validated on astrophysical massive data classification problems, through a web application resource (DAMEWARE), specialized in data mining based on Machine Learning paradigms. Since genetic algorithms are inherently parallel, the GPGPU computing paradigm has provided an exploit of the internal training features of the model, permitting a strong optimization in terms of processing performances and scalability.
\end{abstract}

\section{Introduction}

Computing has started to change how science is done, enabling new scientific advances through enabling new kinds of experiments. They are also generating new kinds of data of increasingly exponential complexity and volume. Achieving the goal of being able to use, exploit and share most effectively these data is a huge challenge. The harder problem for the future is heterogeneity, of platforms, data and applications, rather than simply the scale of the deployed resources. Current platforms require the scientists to overcome computing barriers between them and the data \cite{proceeding1}.\\
The present paper concerns the design and development of a multi-purpose genetic algorithm implemented with the GPGPU/CUDA parallel computing technology. The model comes out from the machine learning supervised paradigm, dealing with both regression and classification scientific problems applied on massive data sets. The model was derived from the original serial implementation, named GAME (Genetic Algorithm Model Experiment) deployed on the DAME \cite{url1} Program hybrid distributed infrastructure and made available through the DAMEWARE \cite{url2} data mining (DM) web application. In such environment the GAME model has been scientifically tested and validated on astrophysical massive data sets problems with successful results \cite{jour1}. As known, genetic algorithms are derived from Darwin's evolution law and are intrinsically parallel in its learning evolution rule and processing data patterns. The parallel computing paradigm can indeed provide an optimal exploit of the internal training features of the model, permitting a strong optimization in terms of processing performances.

\section{Data Mining based on Machine Learning and parallel computing}

Let's start from a real and fundamental assumption: we live in a contemporary world submerged by a tsunami of data. Many kinds of data, tables, images, graphs, observed, simulated, calculated by statistics or acquired by different types of monitoring systems. The recent explosion of World Wide Web and other high performance resources of Information and Communication Technology (ICT) are rapidly contributing to the proliferation of such enormous information repositories.
Machine learning (ML) is a scientific discipline concerned with the design and development of algorithms that allow computers to evolve behaviors based on empirical data. A \emph{learner} can take advantage of examples (data) to capture characteristics of interest of their unknown underlying probability distribution.
These data form the so called Knowledge Base (KB): a sufficiently large set of examples to be used for training of the ML implementation, and to test its performance. The DM methods, however, are also very useful to capture the complexity of small data sets and, therefore, can be effectively used to tackle problems of much smaller scale \cite{jour1}.\\
DM on Massive Data Sets (MDS) poses two important challenges for the computational infrastructure: asynchronous access and scalability. With synchronous operations, all the entities in the chain of command (client, workflow engine, broker, processing services) must remain up for the duration of the activity: if any component stops, the context of the activity is lost.\\
Regarding scalability, whenever there is a large quantity of data, the more affordable approach to making learning feasible relies in splitting the problem in smaller parts (parallelization) sending them to different CPUs and finally combine the results together. So far, the parallel computing technology chosen for this purpose was the GPGPU.\\
GPGPU is an acronym standing for General Purpose Computing on Graphics Processing Units. It was invented by Mark Harris in 2002, \cite{book1}, by recognizing the trend to employ GPU technology for not graphic applications.
With such term we mean all techniques able to develop algorithms extending computer graphics but running on graphic chips. In general the graphic chips, due to their intrinsic nature of multi-core processors (many-core) and being based on hundreds of floating-point specialized processing units, make many algorithms able to obtain higher (one or two orders of magnitude) performances than usual CPUs (Central Processing Units). They are also cheaper, due to the relatively low price of graphic chip components.\\
The choice of graphic device manufacturers, like NVIDIA Corp., was the many-core technology (usually many-core is intended for multi-core systems over 32 cores). The many-core paradigm is based on the growth of execution speed for parallel applications. Began with tens of cores smaller than CPU ones, such kind of architectures reached hundreds of core per chip in a few years. Since 2009 the throughput peak ratio between GPU (many-core) and CPU (multi-core) was about 10:1. Such a large difference has pushed many developers to shift more compu-ting-expensive parts of their programs on the GPUs.

\section{The GAME Model}

An important category of supervised ML models and techniques, in some way related with the Darwin's evolution law, is known as evolutionary (or genetic) algorithms, sometimes also defined as based on genetic programming \cite{book2}.
The slight conceptual difference between evolutionary and genetic algorithms is that the formers are problem-dependent, while the latters are very generic.\\
GAME is a pure genetic algorithm specially designed to solve supervised optimizations problems related with regression and classification functionalities, scalable to efficiently manage MDS and based on the usual genetic evolution methods (crossover, genetic mutation, roulette/ranking, elitism). In order to give a level of abstraction able to make simple to adapt the algorithm to the specific problem, a family of polynomial developments was chosen for GAME model. This methodology makes the algorithm itself easily expandable, but this abstraction requires a set of parameters that allows fitting the algorithm to the specific problem.\\
From an analytic point of view, a pattern, composed of N features contains an amount of information correlated between the features corresponding to the target value. Usually in a real scientific problem that correlation is \emph{masked} from the noise (both intrinsic to the phenomenon, and due to the acquisition system); but the unknown correlation function can ever be approximated with a polynomial sequence, in which the degree and non-linearity of the chosen function determine the approximation level. The generic function of a polynomial sequence is based on these simple considerations:\\
Given a generic dataset with N features and a target t, pat a generic input pattern of the dataset, $pat = (f_1, ..., f_N, t)$ and $g(x)$ a generic real function, the representation of a generic feature $f_i$ of a generic pattern, with a polynomial sequence of degree d is:

\begin{equation}
G({f_i}) \cong a_0 + a_1g({f_i}) + ... + a_dg^d(f_i )
\end{equation}

Hence, the k-th pattern $(pat_k)$ with N features may be represented by:

\begin{equation}
Out({pat_k}) \cong \sum\limits_{i = 1}^N {G({f_i}) \cong a_0 + \sum\limits_{i = 1}^N {\sum\limits_{j = 1}^d {a_j g^j({f_i})} } }
\end{equation}

Then target $t_k$, concerning to pattern $pat_k$, can be used to evaluate the approximation error of the input pattern to the expected value:

\begin{equation}
E_k  = ({t_k - Out({pat_k})})^2
\end{equation}

If we generalize the expression (2) to an entire dataset, with NP number of patterns $(k = 1,..., NP)$, at the end of the \emph{forward} phase (batch) of the GA, we obtain NP expressions (2) which represent the polynomial approximation of the dataset.\\
In order to evaluate the fitness of the patterns as extension of (3), the Mean Square Error (MSE) or Root Mean Square Error (RMSE) may be used.\\
Then we define a GA with the following characteristics:

\begin{itemize}
\item The expression (2) is the fitness function;
\item The array $(a_0,..., a_M)$ defines M genes of the generic chromosome (initially they are generated random and normalized between -1 and +1);
\item All the chromosomes have the same size (constrain from a classic GA);
\item The expression (3) gives the standard error to evaluate the fitness level of the chromosomes;
\item The population (genome) is composed by a number of chromosomes imposed from the choice of the function $g(x)$ of the polynomial sequence.
\end{itemize}

About the last item, this number is determined by the following expression:

\begin{equation}
NUM_{chromosomes}  = (B \cdot N) + 1
\end{equation}

where N is the number of features of the patterns and B is a multiplicative factor that depends from the $g(x)$ function, which in the simplest case is just 1, but can arise to 3 or 4 in more complex cases. The parameter B also influences the dimension of each chromosome (number of genes):

\begin{equation}
NUM_{genes}  = (B \cdot d) + 1
\end{equation}

where d is the degree of the polynomial. For example if we use the trigonometric polynomial expansion, given by the following expression (hereinafter polytrigo),

\begin{equation}
g(x) = a_0  + \sum\limits_{m = 1}^d {a_m \cos (mx) + } \sum\limits_{m = 1}^d {b_m \sin (mx)}
\end{equation}

in order to have 200 patterns composed by 11 features, the expression using (2) with degree 3, will become:

\begin{equation}
Out(pat_{k = 1...200} ) \cong \sum\limits_{i = 1}^{11} {G(f_i ) \cong a_0  + \sum\limits_{i = 1}^{11} {\sum\limits_{j = 1}^3 {a_j \cos (jf_i ) + } } \sum\limits_{i = 1}^{11} {\sum\limits_{j = 1}^3 {b_j \sin (jf_i )} } }
\end{equation}

In the last expression we have two groups of coefficients (sin and cosine), so B will assume the value 2. Hence the generic genome (population at a generic evolution stage), will be composed by 23 chromosomes, given by equation (4), each one with 7 genes $[a_0, a_1, a_2, a_3, b_1, b_2, b_3]$, given by equation (5), with each single gene (coefficient of the polynomial) in the range $[-1, +1]$.\\
In the present project, the idea is to build a GA able to solve supervised crispy classification and regression problems, typically related to an high-complexity parameter space where the background analytic function is not known, except for a limited number of couples of input-target values, representing valid solutions to a physical category of phenomena. A typical case is to classify astronomical objects based on some solution samples (the KB) or to predict new values extracted by further observations. To accomplish such behavior we designed a function (a polynomial expansion) to combine input patterns. The coefficients of such polynomials are the chromosome genes. The goal is indeed to find the best chromosome so that the related polynomial expansion is able to approximate the right solutions to input pattern classification/regression. So far, the fitness function for such representation consists of the training error, obtained as absolute difference between the polynomial output and the target value for each pattern. Due to the fact that we are interested to find the minimum value of the error, the fitness is calculated as the complement of the error (i.e. 1-error) and the problem is reduced to find the chromosome achieving the maximum value of fitness.\\

\section{The GPU-based GAME implementation}

In all execution modes (use case), GAME exploits the polytrigo function (6), consisting in a polynomial expansion in terms of sum of sins and cosines. Specifically in the training use case, corresponding to the GA building and consolidation phase, the polytrigo is used at each iteration as the transformation function applied to each chromosome to obtain the output on the problem input dataset, and indirectly also to evaluate the fitness of each chromosome. It is indeed one of the critical aspects of the serial algorithm to be investigated during the parallelization design process.\\
Moreover, after having calculated the fitness function for all genetic population chromosomes, this information must be back-propagated to evolve the genetic population. This back and forth procedure must be replicated as many times as it is the training iteration number or the learning error threshold, both decided and imposed by the user at setup time of any experiment. The direct consequence of the above issues is that the training use case takes much more execution time than the others (such as test and validation), and therefore is the one we are going to optimize.\\
Main design aspect approaching the software architecture analysis for the GPU is the partition of work: i.e. which work should be done on the CPU vs. the GPU. We have identified the time consuming critical parts to be parallelized by executing them on the GPU. They are the generation of random chromosomes and the calculation of the fitness function of chromosomes.
The key principle is that we need to perform the same instruction simultaneously on as much data as possible. By adding the number of chromosomes to be randomly generated in the initial population as well as during each generation, the total number of involved elements is never extremely large but it may occur with a high frequency. This is because also during the population evolution loop a variable number of chromosomes are randomly generated to replace older individuals. To overcome this problem we may generate a large number of chromosomes randomly \emph{una tantum}, by using them whenever required. On the contrary, the evaluation of fitness functions involves all the input data, which is assumed to be massive datasets, so it already has an intrinsic data-parallelism.
Since CUDA programming involves code running concurrently on a host with one or more CPUs and one or more CUDA-enabled GPU, it is important to keep in mind that the differences between these two architectures may affect application performance to use CUDA effectively. The function polytrigo takes about three-quarters of the total execution time of the application, while the total including child functions amounts to about 7/8 of total time execution. This indeed has been our first candidate for parallelization. In order to give a practical example, for the interested reader, we report the source code portions related to the different implementation of the polytrigo function, of the serial and parallelized cases.

\medskip

\noindent
{\it C++ serial code for polytrigo function (equation 6):}
\begin{verbatim}

for (int i = 0; i < num_features; i++) {
   for (int j = 1; j <= poly_degree; j++) {
	ret +=  v[j] * cos(j * input[i]) + v[j + poly_degree] *
            * sin(j * input[i]); } }
\end{verbatim}

\medskip

\noindent
{\it CUDA C (Thrust) parallelized code for polytrigo function (equation 6):}
\begin{verbatim}

struct sinFunctor {  __host__ __device__
 double operator()(tuple <double, double> t) {
    return sin(get < 0 > (t) * get < 1 > (t)); }};
struct cosFunctor {  __host__ __device__
 double operator()(tuple <double, double> t) {
    return cos(get < 0 > (t) * get < 1 > (t)); }};
thrust::transform(thrust::make_zip_iterator(
    thrust::make_tuple(j.begin(), input.begin())),
    thrust::make_zip_iterator(
       thrust::make_tuple(j.end(), input.end())),
       ret.begin(), sinFunctor(), cosFunctor());
\end{verbatim}

Noting that, while the vector $v[]$ is continuously evolving, $input[]$ (i.e. the elements of the input dataset) are being used in calculation of ret at each iteration but they are never altered. We rewrite the function by calculating in advance the sums of sins and cosines, storing the results in two vectors and then use them in the function polyTrigo() at each iteration.
This brings huge benefits because we calculate trigonometric functions, which are those time consuming, only once instead of at every iteration and exploit the parallelism on large amount of data because it assumes that we have large input datasets.\\
From the time complexity point of view, by assuming to have as many GPU cores as population chromosomes, the above CUDA C code portion would take constant time, instead of polynomial time required by the corresponding C++ serial code.

\section{The Experiment}

In terms of experiments, the two CPU versions of GAME, the original and an optimized version of the serial algorithm (hereinafter serial and Opt respectively), together with the final version for GPU (hereinafter ELGA), have been compared basically by measuring their performance in terms of execution speed, by also performing an intrinsic evaluation of the overall scientific performances. The optimized algorithm is the serial version adapted by modifying the code portions which are candidate to be parallelized in the final GPU release.\\
Initially, the tests have been organized by distinguishing between classification and regression functional modes. By analyzing early trials, however, it resulted that the performance growth was virtually achieved in both cases. So far, we limit here the discussion details to a classification experiment, done in the astrophysical context.\\
The scientific problem used here as a test bed for data mining application of the GAME model is the search (classification) of Globular Cluster (GC) populations in external galaxies \cite{jour1}. This topic is of interest to many astrophysical fields: from cosmology, to the evolution of stellar systems, to the formation and evolution of binary systems.\\
The dataset used in this experiment consists in wide field HST observations of the giant elliptical NGC1399 in the Fornax cluster, \cite{jour2}. The subsample of sources used to build our Base of Knowledge, to train the GAME model is composed by  2100 sources with all photometric and morphological information, \cite{jour1}. Finally, our classification dataset consisted of 2100 patterns, each composed by 11 features (including the two targets, corresponding to the classes GC and not GC used during the supervised training phase).\\
The performance was evaluated on several hardware platforms. We compared our production GPU code with a CPU implementation of the same algorithm. The benchmarks were run on a 2.0 GHz Intel Core i7 2630QM quad core CPU running 64-bit Windows 7 Home Premium SP1. The CPU code was compiled using the Microsoft® C/C++ Optimizing Compiler version 16.00 and GPU benchmarks were performed using the NVIDIA CUDA programming toolkit version 4.1 running on several generations of NVIDIA GPUs GeForce GT540M.
As execution parameters were chosen combinations of:

\begin{itemize}
\item Max number of iterations: 1000, 2000, 4000, 10000, 20000 and 40000;
\item Order (max degree) of polynomial expansion: 1, 2, 4 and 8;
\end{itemize}

The other parameters remain unchanged for all tests:

\begin{itemize}
\item error function (fitness): MSE with threshold = 0.001;
\item Selection criterion: RANKING and ROULETTE;
\item Crossover probability rate 0.9 and mutation probability rate 0.2;
\item Elitism chromosomes at each evolution: 2.
\end{itemize}

For the scope of the present experiment, we have preliminarily verified the perfect correspondence between CPU- and GPU-based implementations in terms of classification performances. In fact, the scientific results for the CPU-based algorithm have been already evaluated and documented in a recent paper \cite{jour1}, where the CPU version of GAME were also compared with other ML models, provided by our team.\\
Referring to the best results as described in \cite{jour1}, for the serial code version we obtained the following percentages on a dataset consisting of 2100 patterns each composed by 7 column features:

\begin{itemize}
\item Classification accuracy = $86.4\%$;
\item Completeness = $78.9\%$;
\item Contamination = $13.9\%$;
\end{itemize}

\begin{figure}
\centering
\includegraphics[height=6.2cm]{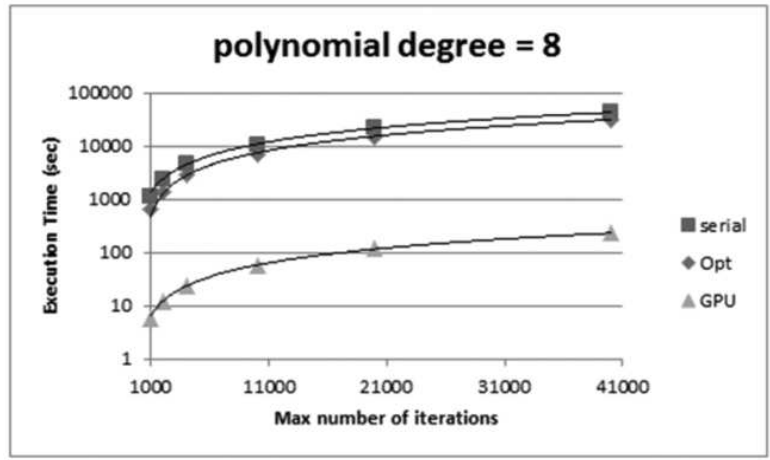}
\caption{Comparison among the GAME implementations with the polynomial degree = 8.}
\label{fig:1}
\end{figure}

In both optimized serial and parallelized version, we obtained, as expected, the same values, slightly varying in terms of least significant digit, trivially motivated by the intrinsic randomness of the genetic algorithms.\\
Here we investigated the analysis of performances in terms of execution speed. By using the defined metrics we compared the three versions of GAME implementation, under the same setup conditions. As expected, while the two CPU-based versions, serial and Opt, appear comparable, there is a quite shocking difference with the GPU-based version ELGA. The diagram of Fig.~\ref{fig:1} reports the direct comparisons among the three GAME versions, by setting a relatively high degree of the polynomial expansion which represents the evaluation function for chromosomes.

We performed also other tests, by varying the polynomial degree. The trends show that the execution time increases always in a linear way with the number of iterations, once fixed the polynomial degree. This is what we expected because the algorithm repeats the same operations at each iteration. The GPU-based version speed is always at least one order of magnitude less than the other two implementations. We remark also that the classification performances of the GAME model increases by growing the polynomial degree, starting to reach good results from a value equal to 4. Exactly when the difference between CPU and GPU versions starts to be 2 orders of magnitude.

\begin{figure}
\centering
\includegraphics[height=6.2cm]{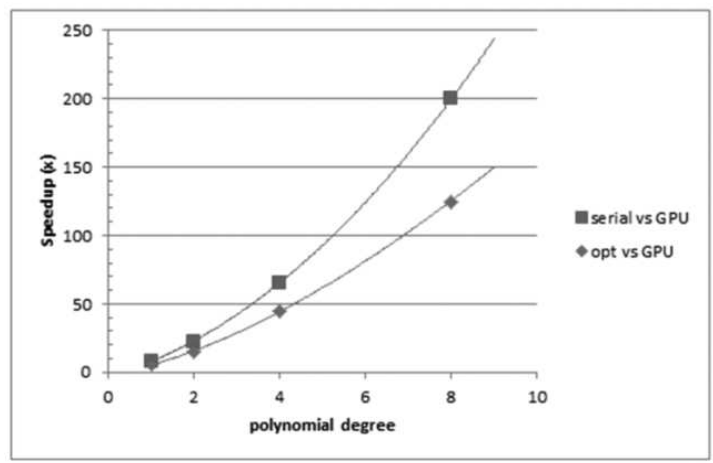}
\caption{Speedup comparison among GAME CPU implementations against the GPU version.}
\label{fig:2}
\end{figure}

In the diagram of Fig.~\ref{fig:2}, the GPU version is compared against the CPU implementations. As shown, the speedup increases proportionally with the increasing of the polynomial degree. The diagram shows that for the average speed in a range of iterations from 1000 to 40000, the ELGA algorithm exploits the data parallelism as much data are simultaneously processed. As previously mentioned, an increase of maximum degree in the polynomial expansion leads to an increase in the number of genes and consequently to a larger population matrix. The GPU algorithm outperforms the CPU performance by a factor ranging from $8x$ to $200x$ in the not optimized (serial) case and in a range from $6x$ to $125x$ in the optimized case (Opt), enabling an intensive and highly scalable use of the algorithm that were previously impossible to be achieved with a CPU.

\section{Conclusions}

We investigated the state of the art computing technologies, by choosing the one best suited to deal with a wide range of real physical problems. A multi-purpose genetic algorithm (GA) implemented with GPGPU/CUDA parallel computing technology has been designed and developed. The model comes from the paradigm of supervised machine learning, addressing both the problems of classification and regression applied on massive data sets.\\
The model was derived from a serial implementation named GAME, deployed on the DAME Program, \cite{proceeding2} and \cite{proceeding3}, hybrid distributed infrastructure and already scientifically tested and validated on astrophysics massive data sets problems with successful results. Since GAs are inherently parallel, the parallel computing paradigm has provided an exploit of the internal training features of the model, permitting a strong optimization in terms of processing performances. We described our effort to adapt our genetic algorithm for general purpose on GPU. We discussed the efficiency and computational costs of various components involved that are present in the algorithm. Several benchmark results were shown. The use of CUDA translates into a $75x$ average speedup, by successfully eliminating the largest bottleneck in the multi-core CPU code. Although a speedup of up to $200x$ over a modern CPU is impressive, it ignores the larger picture of use a Genetic Algorithm as a whole. In any real-world the dataset can be very large (those we have previously called Massive Data Sets) and this requires greater attention to GPU memory management, in terms of scheduling and data transfers host-to-device and \emph{vice versa}. Moreover, the identical results for classification functional cases demonstrate the consistency of the implementation for the three different computing architectures, enhancing the scalability of the proposed GAME model when approaching massive data sets problems.\\
Finally, the very encouraging results suggest to investigate further optimizations, like: (i) moving the formation of the population matrix and its evolution in place on the GPU. This approach has the potential to significantly reduce the number of operations in the core computation, but at the cost of higher memory usage; (ii) exploring more improvements by mixing Thrust and CUDA C code, that should allow a modest speedup justifying development efforts at a lower level; (iii) use of new features now available on NVIDIA Fermi architecture, such as faster atomics and more robust thread synchronization and multi GPUs capability.

\subsubsection*{Acknowledgments.} The experiment has been done by means a M.Sc. degree  in Informatics Engineering arisen from a Collaboration among several Italian academic institutions. The hardware resources have been provided by Dept. of Computing Engineering and Systems and S.Co.P.E. GRID Project infrastructure of the University Federico II of Naples. The data mining model has been designed and developed by DAME Program Collaboration. MB wishes to thank the financial support of PRIN-INAF 2010, "Architecture and Tomography of Galaxy Clusters". This work has been partially funded by LINCE project of the F.A.R.O. programme jointly financed by the Compagnia di San Paolo and by the Polo delle Scienze e delle Tecnologie of the University of Napoli "Federico II" and it has been carried out also thanks to a hardware donation in the context of the NVIDIA Academic Partnership program. The authors also wish to thank the financial support of Project F.A.R.O. III Tornata, from University Federico II of Naples.

\end{document}